\documentclass[11pt]{article}
\pdfoutput=1
\usepackage[ngerman]{babel}
\usepackage[a4paper,text={16.8cm,22.4cm}]{geometry}
\usepackage{amsmath,amsfonts,braket,slashed,amssymb,tikz,bm,psfrag,graphicx,color,dsfont}
\usepackage{multicol}
\usepackage{ctable}
\usepackage[small,labelfont=bf]{caption}
\usepackage{hyperref}
\usepackage{subcaption}
\usepackage{cleveref}
\usepackage{bbm, soul, nicefrac,graphicx,wrapfig, setspace}

\sodef\an{}{.15em}{0.5em plus0.2em}{0.5em plus.1em minus.1em}

\begin{document}
\begin{center}
\Large\bf\boldmath
The Stern-Gerlach Experiment\\
Translation of: ``Der experimentelle Nachweis der Richtungsquantelung im Magnetfeld" \,\end{center}
\begin{center}
Martin Bauer\\
Institute for Particle Physics Phenomenology, Department of Physics,\\
Durham University, Durham, DH1 3LE, United Kingdom\\[2mm]
\end{center}

The following is a translation of the paper by Walther Gerlach and Otto Stern\footnote{W. Gerlach u. O. Stern, Zeitschrift f{\"u}r Physik \textbf{9}, 349--352, 1922.}) that reported the first evidence for the quantisation of atoms in a magnetic field. The atoms have quantum states corresponding to a limited number of possible angles between the directions of the angular momenta of the atoms and the magnetic field, also called space quantisation. Wording and layout have been chosen to be as close to the original as possible. For context we recommend the recent review\footnote{H. Schmidt-B{\"o}cking, L.~Schmidt, H.~L{\"u}dde,  W.~Trageser, A.~Templeton and T.~Sauer, Eur. Phys. J. H \textbf{41}, 327--364, 2016.}). \\
I thank Phillip Helbig for substantial help in preparing this translation and Chanda Prescod-Weinstein for bringing to my attention that there is no available english translation of the original paper by Gerlach and Stern.

\newpage

\vspace{0.5cm}
\begin{center}
\Large\bf\boldmath
Experimental Evidence for Space Quantisation in a Magnetic Field.
\end{center}
\begin{center}By  \textbf{Walther Gerlach} in Frankfurt a.M.  and \textbf{Otto Stern} in Rostock. \end{center}
\begin{center}Including 7 figures. (Recieved March 1, 1922.)
\end{center}
 
Recently\footnote{O. Stern, ZS. f. Phys. \textbf{7}, 249, 1921}) this journal published a possible method to experimentally answer the question whether space quantisation in a magnetic field exists. It was shown in a second communication\footnote{W. Gerlach u. O. Stern, ibid. \textbf{8}, 110, 1921.}) that the normal silver atom has a magnetic moment. We allow ourselves to report in the following that the continuation of these investigations has led to \an{establish space quantisation in a magnetic field as a fact.}\\
\an{Experimental setup.} Method and experimental setup are generally the same as in our earlier experiments, but substantial improvements
have been made to some parts of it\footnote{These were worked out and tested collaboratively. The final experiments had to be performed by one of us alone (G.) due to the departure from Frankfurt of one of us (St.). }), which we will describe here in addition to the  \begin{wrapfigure}{r}{2.5cm}
\includegraphics[width=3cm]{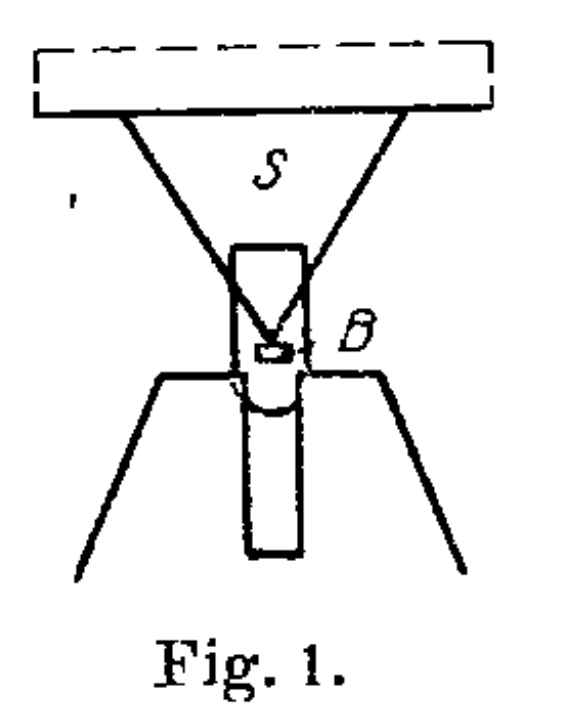}
\end{wrapfigure}
 information provided earlier.   The beam of silver atoms emerges from an electrically heated small chamotte oven with a steel insert and a cover with a 1 mm$^2$ circular hole. The distance between the hole in the oven and the first beam aperture was increased to 2.5 cm to prevent clogging by occasional small silver droplets sprayed from the oven as well as precipation from the atomic beam. This first aperture is almost circular and its surface measures $3\cdot 10^{-3}$ mm$^2$. 3.3 cm behind this circular aperture, the silver beam passes through a slit aperture with a length of 0.8 mm and a width between 0.03 mm and 0.04 mm. Both apertures are made from platinum sheet. The slit aperture is positioned where the magnetic field begins. The opening of the slit aperture is right above the knife-edge $S$ (see Fig. 1) and is adjusted with respect to the circular aperture and the hole in the oven in such a way that the silver beam travels in parallel along the 3.5 cm long knife-edge. Precisely at the end of this knife-edge the silver beam hits a glass plate where it condenses.

The two apertures, the poles of the magnet and the glass plate are in a brass housing with wall of thickness 1 cm, which are rigidly connected so that pressure from the poles of the electromagnet doesn't result in a deformation of the housing nor cause a shift in the relative position of the apertures, the poles, and the glass plate.

Two Volmer diffusion pumps and a Gaede Hg-pump as pre-pump are used for evacuation. Through continuous pumping and cooling with solid carbon dioxide a vacuum of $10^{-5}$mm Hg was achieved and maintained.

\begin{figure}[h!]
\hspace{-1cm}\includegraphics[width=1.2\textwidth]{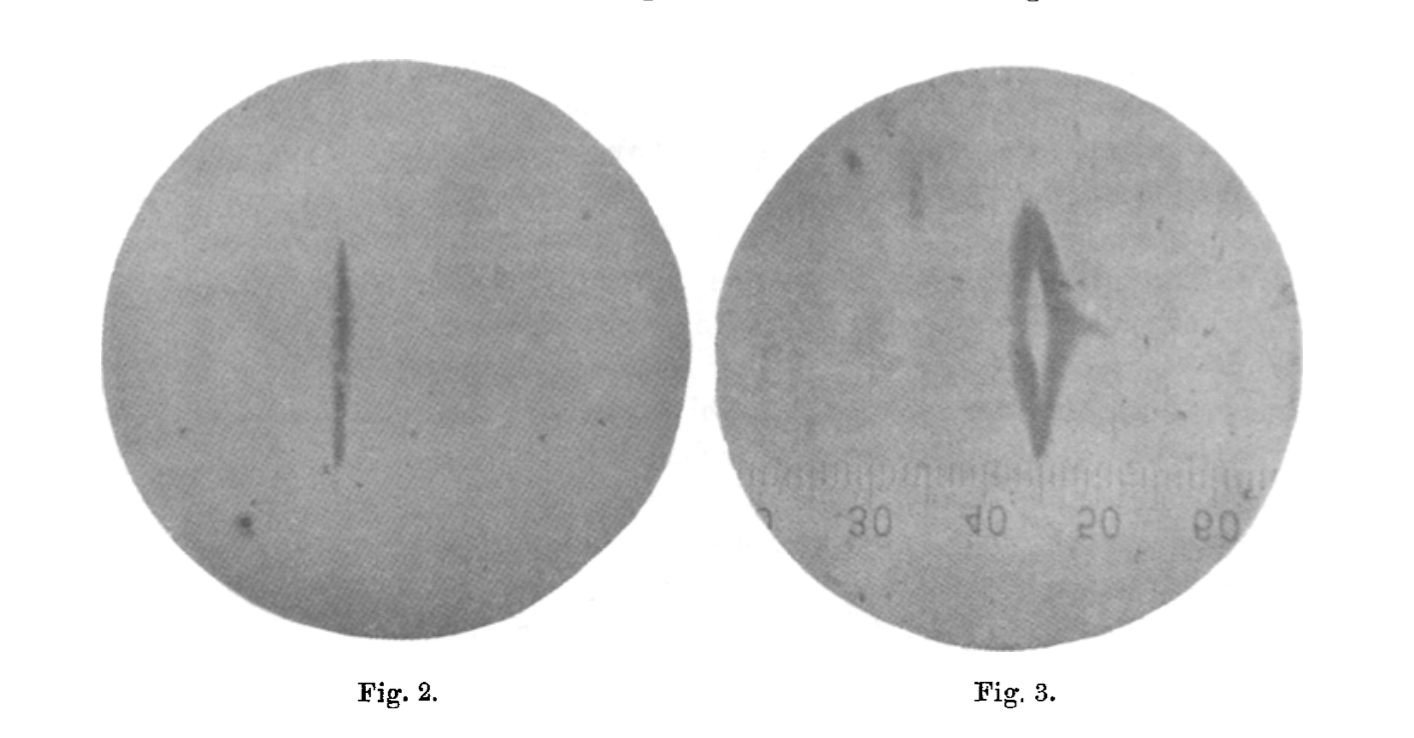}
\end{figure}

The ``exposure time"\,was increased to eight hours without interruption. But as a result of the very narrow apertures and the long beam, the silver film on the glass plate was so thin that it had to be developed ---as reported previously--- even after eight hours of vaporisation.\\
\an{Results.} First, Fig. 2 shows a photograph after 4\nicefrac{1}{2} hours of exposure time without a magnetic field; it's magnified by a factor of 20. Measurements of the original under the microscope using an ocular micrometer resulted in the following dimensions: Length 1.1 mm, width at the narrowest point 0.06 mm, at the widest point 0.10 mm.  We see that the slit isn't exactly parallel. It should be noted, however, that the figure shows the slit magnified by a factor of 40, since the ``silver image"\,of the slit is already twice the size; it is difficult to make such a slit in a frame only a few millimetres in size. 
\newpage
Fig. 3 shows a photograph after eight hours of exposure time magnified by a factor 20 (20 scale divisions of the imaged scale = 1 mm). This is the best photograph we took. Two other photographs show the same result in all relevant aspects, but don't show this complete symmetry. It has to be said that adjustment of these small apertures by optical methods is very difficult, so that it takes some luck to obtain a perfectly symmetric photograph as shown in Fig.3; errors in adjusting an aperture by just a few hundredths of a millimeter are enough to completely ruin a photograph.

\begin{figure}[h!]
\hspace{-.5cm}\includegraphics[width=1.1\textwidth]{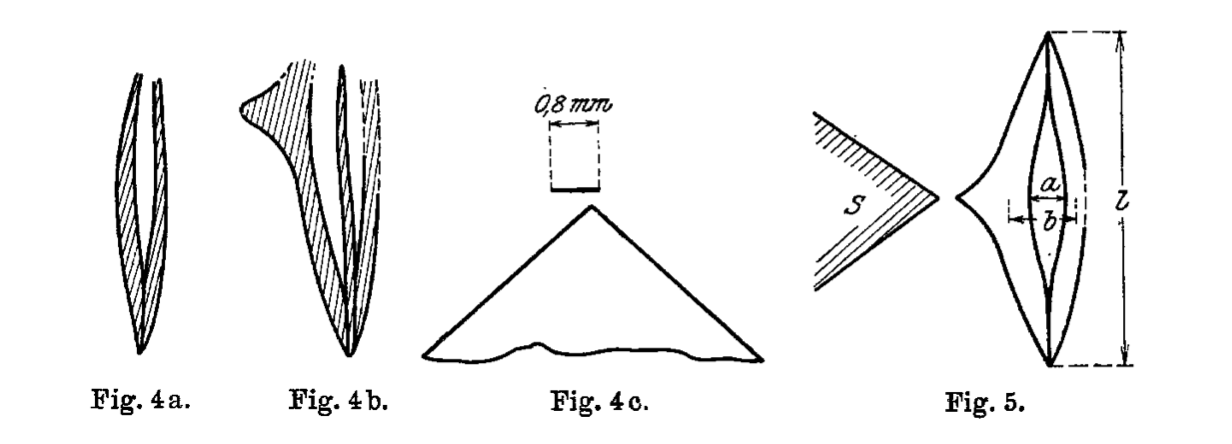}
\end{figure}

The results of the two other experiments are shown schematically in Fig. 4a and 4b. In Fig. 4a the silver beam ran intentionally at a slightly greater distance past the knife-edge than in the experiment of Fig.3. The slit aperture wasn't completely ``filled" \,here. 
Fig. 4b shows the deposit of an experiment both with and without a field on the same plate; the beam ran very close to the knife-edge, but was moved 0.3 mm perpendicular to the field (Fig. 4c). Regarding clarity of the pictures, the complete splitting, and all other details, these pictures are in no way inferior to those shown in Fig. 3. 

The photographs show that the silver-atom beam is split in an inhomogeneous magnetic field in two directions in the direction of the inhomogeneity, of which one is attracted by the knife-edge pole and the other is repelled by the knife-edge pole. The deposits show the following details (see the schematic Fig. 5)

\begin{itemize}
\item[a)] \an{The dimensions} of the original were measured in a microscope: Length 1.1 mm, width $a$ 0.11 mm, width $b$ 0.20 mm.
\item[b)] \an{The atomic beam splits into two discrete beams in a magnetic field. We found no evidence for non-deflected atoms. }
\item[c)] \an{The attraction is slightly stronger than the repulsion.} The attracted atoms get closer to the pole and therefore to the region of largest inhomogeneity, so that the deflection while flying past is stronger. Fig. 3 and 4b show the significantly larger deflection directly at the knife-edge of this one magnetic pole. In the immediate vicinity of the knife-edge the attraction becomes very large so that a bulge arises, with a sharp edge pointing towards the knife-edge. 
\item[d)] \an{The width of the deflected bands is larger than the width of the undeflected image.} The latter is simply the image of aperture $B_2$ projected by aperture $B_1$ onto the glass plate. The deflected band is broadened following the velocity distribution of the silver atoms. 
\item[e)] \an{This fact strengthens the case for there being few if any undeflected atoms [see b)].} 
The detection of undeflected atoms in a small area is much more sensitive than that of deflected atoms in a large area.  There appears to be no magnetic axis perpendicular to the field direction.
\end{itemize}

\an{We view these results as direct, experimental evidence for space quantisation in a magnetic field.}

A detailed account of the experiment and results in our thus far short communications will be published in the Annalen der Physik, as soon as we have precise measurements of the inhomogeneity of the magnetic field and can provide quantitative information about the size of the magneton.

The electromagnet necessary for these experiments was procured with funds from the foundation of the Kaiser Wilhelm Institute; to the director, Mr. A. Einstein, we would like to express our heartfelt thanks. We further thank the Association of Friends and Sponsors of the University of Frankfurt sincerely for the abundant resources they gladly made available to fund the continuation of the experiments.\\[2pt]

Frankfurt a. M. and Rostock i. M., February 1922.

   \begin{center}
   $\rule{3cm}{0.04cm}$
     \end{center}

\end{document}